\documentclass[a4paper,12pt]{article}
\usepackage{CJKutf8} 
\usepackage{jheppub}
\usepackage{expl3}
\usepackage{amsmath,amssymb,mathrsfs,amsthm,tikz,shuffle,paralist}
\usepackage{dsfont}
\usepackage{color}
\usepackage{enumitem}
\definecolor{darkred}{RGB}{173,34,48}
\definecolor{Nblue}{RGB}{0,47,167}   
\definecolor{ku}{RGB}{144,26,30}
\usepackage{tcolorbox}

\usepackage{url}

\usepackage{breakurl}
\usepackage{hyperref}
\hypersetup{breaklinks=true}

\usetikzlibrary{shapes.misc}
\usetikzlibrary{cd}
\usetikzlibrary{snakes}

\usepackage[all]{xy}

\makeatletter
\DeclareRobustCommand*{\bfseries}{%
  \not@math@alphabet\bfseries\mathbf
  \fontseries\bfdefault\selectfont
  \boldmath
}

\usepackage[textsize=tiny]{todonotes}

\newcommand{\Maple}{\texttt{\textup{Maple}}}
\newcommand{\HyperInt}{\texttt{\textup{HyperInt}}}

\newcommand{\Ef}[3]{{\textrm{E}_4}(\begin{smallmatrix}#1\\#2%
\end{smallmatrix};#3)} 
\newcommand{\cEf}[3]{{\mathcal{E}_4}(\begin{smallmatrix}#1\\#2%
\end{smallmatrix};#3)} 
\newcommand{\gamt}[3]{{\widetilde{\Gamma}}(\begin{smallmatrix}#1\\#2%
\end{smallmatrix};#3)}

\newcommand{\drawelliptic}{%
\begin{tikzpicture}[scale=0.2, label distance=-1mm,baseline={([yshift=-.5ex]current bounding box.center)}]
\node (v20) at (-3.5,2.5) {};
\node (v26) at (-1.5,2.5) {};
\node (v27) at (0.5,2.5) {};
\node at (2.5,2.5) {};
\node (v35) at (4.5,2.5) {};
\node (v39) at (6.5,2.5) {};
\node (v42) at (8.5,2.5) {};
\node (v23) at (-3.5,0) {};
\node (v25) at (-1.5,0) {};
\node (v28) at (0.5,0) {};
\node (v38) at (2.5,0) {};
\node (v36) at (4.5,0) {};
\node (v40) at (6.5,0) {};
\node (v41) at (8.5,0) {};
\node (v19) at (-5,3) {};
\node (v21) at (-4,4) {};
\node (v22) at (-5,-0.5) {};
\node (v24) at (-4,-1.5) {};
\node (v45) at (10,-0.5) {};
\node (v44) at (10,3) {};
\node (v43) at (9,4) {};
\node (v46) at (9,-1.5) {};
\node (v37) at (2.5,4) {};
\node (v1) at (-2.5,1.25) {};
\node (v12) at (-1.5,1.25) {};
\node (v13) at (0.5,1.25) {};
\node (v4) at (1.5,1.25) {};
\node (v9) at (3.5,1.25) {};
\node (v16) at (4.5,1.25) {};
\node (v17) at (6.5,1.25) {};
\node (v7) at (7.5,1.25) {};
\node (v47) at (-2.5,2.5) {};
\node (v48) at (-2.5,0) {};
\node (v52) at (1.5,0) {};
\node (v49) at (1.5,2.5) {};
\node (v55) at (3.5,2.5) {};
\node (v56) at (3.5,0) {};
\node (v58) at (7.5,0) {};
\node (v57) at (7.5,2.5) {};
\node (v60) at (-0.5,2.5) {};
\node (v59) at (-0.5,0) {};
\node (v63) at (5.5,0) {};
\node (v64) at (5.5,2.5) {};
\node at (-0.5,1.25) {};
\node at (2.5,1.25) {};
\node at (5.5,1.25) {};
\node (v5) at (2.5,-1.5) {};
\draw[thick]  (v19.center) edge (v20.center);
\draw[thick]  (v20.center) edge (v21.center);
\draw[thick]  (v20.center) edge (v23.center);
\draw[thick]  (v23.center) edge (v22.center);
\draw[thick]  (v23.center) edge (v24.center);
\draw[thick]  (v23.center) edge (v25.center);
\draw[thick]  (v20.center) edge (v26.center);
\draw[thick]  (v26.center) edge (v25.center);
\draw[thick]  (v28.center) edge (v27.center);
\draw[thick]  (v27.center) edge (v35.center);
\draw[thick]  (v35.center) edge (v36.center);
\draw[thick]  (v36.center) edge (v28.center);
\draw[thick]  (v38.center) edge (v37.center);
\draw[thick]  (v38.center) edge (v5.center);
\draw[thick]  (v39.center) edge (v40.center);
\draw[thick]  (v40.center) edge (v41.center);
\draw[thick]  (v41.center) edge (v42.center);
\draw[thick]  (v39.center) edge (v42.center);
\draw[thick]  (v42.center) edge (v43.center);
\draw[thick]  (v42.center) edge (v44.center);
\draw[thick]  (v41.center) edge (v45.center);
\draw[thick]  (v41.center) edge (v46.center);
\draw[thick]  (v26.center) edge (v27.center);
\draw[thick]  (v25.center) edge (v28.center);
\draw[thick]  (v35.center) edge (v39.center);
\draw[thick]  (v36.center) edge (v40.center);
\draw[fill=black] (-0.5,1.25) circle (2pt);
\draw[fill=black] (-0.1,1.25) circle (2pt);
\draw[fill=black] (-0.9,1.25) circle (2pt);
\draw[fill=black] (5.5,1.25) circle (2pt);
\draw[fill=black] (5.1,1.25) circle (2pt);
\draw[fill=black] (5.9,1.25) circle (2pt);
\end{tikzpicture}%
}

\newcommand{\drawstaggered}{%
\begin{tikzpicture}[scale=0.2, label distance=-1mm,baseline={([yshift=-.5ex]current bounding box.center)}]
\node (v20) at (-3.5,2.5) {};
\node (v26) at (-1.5,2.5) {};
\node (v27) at (0.5,2.5) {};
\node (v31) at (2.5,2.5) {};
\node (v29) at (4.5,2.5) {};
\node (v33) at (6.5,2.5) {};
\node at (8.5,2.5) {};
\node (v35) at (10.5,2.5) {};
\node (v39) at (12.5,2.5) {};
\node (v42) at (14.5,2.5) {};
\node (v23) at (-3.5,0) {};
\node (v25) at (-1.5,0) {};
\node (v28) at (0.5,0) {};
\node at (2.5,0) {};
\node (v30) at (4.5,0) {};
\node (v34) at (6.5,0) {};
\node (v38) at (8.5,0) {};
\node (v36) at (10.5,0) {};
\node (v40) at (12.5,0) {};
\node (v41) at (14.5,0) {};
\node (v19) at (-5,3) {};
\node (v21) at (-4,4) {};
\node (v22) at (-5,-0.5) {};
\node (v24) at (-4,-1.5) {};
\node (v45) at (16,-0.5) {};
\node (v44) at (16,3) {};
\node (v43) at (15,4) {};
\node (v46) at (15,-1.5) {};
\node (v37) at (8.5,4) {};
\node (v32) at (2.5,-1.5) {};
\node (v1) at (-2.5,1.25) {};
\node (v12) at (-1.5,1.25) {};
\node (v13) at (0.5,1.25) {};
\node (v4) at (1.5,1.25) {};
\node (v5) at (3.5,1.25) {};
\node (v14) at (4.5,1.25) {};
\node (v15) at (6.5,1.25) {};
\node (v6) at (7.5,1.25) {};
\node (v9) at (9.5,1.25) {};
\node (v16) at (10.5,1.25) {};
\node (v17) at (12.5,1.25) {};
\node (v7) at (13.5,1.25) {};
\draw[thick]  (v19.center) to (v20.center);
\draw[thick]  (v20.center) to (v21.center);
\draw[thick]  (v22.center) to (v23.center);
\draw[thick]  (v23.center) to (v24.center);
\draw[thick]  (v20.center) to (v23.center);
\draw[thick]  (v23.center) to (v25.center);
\draw[thick]  (v20.center) to (v26.center);
\draw[thick]  (v26.center) to (v25.center);
\draw[thick]  (v26.center) to (v27.center);
\draw[thick]  (v25.center) to (v28.center);
\draw[thick]  (v27.center) to (v29.center);
\draw[thick]  (v28.center) to (v30.center);
\draw[thick]  (v27.center) to (v28.center);
\draw[thick]  (v31.center) to (v32.center);
\draw[thick]  (v29.center) to (v30.center);
\draw[thick]  (v29.center) to (v33.center);
\draw[thick]  (v30.center) to (v34.center);
\draw[thick]  (v34.center) to (v33.center);
\draw[thick]  (v33.center) to (v35.center);
\draw[thick]  (v34.center) to (v36.center);
\draw[thick]  (v37.center) to (v38.center);
\draw[thick]  (v35.center) to (v39.center);
\draw[thick]  (v36.center) to (v40.center);
\draw[thick]  (v35.center) to (v36.center);
\draw[thick]  (v39.center) to (v40.center);
\draw[thick]  (v40.center) to (v41.center);
\draw[thick]  (v41.center) to (v42.center);
\draw[thick]  (v42.center) to (v39.center);
\draw[thick]  (v43.center) to (v42.center);
\draw[thick]  (v42.center) to (v44.center);
\draw[thick]  (v41.center) to (v45.center);
\draw[thick]  (v41.center) to (v46.center);
\node (v47) at (-2.5,2.5) {};
\node (v48) at (-2.5,0) {};
\node (v52) at (1.5,0) {};
\node (v49) at (1.5,2.5) {};
\node (v50) at (3.5,2.5) {};
\node (v54) at (3.5,0) {};
\node (v53) at (7.5,0) {};
\node (v51) at (7.5,2.5) {};
\node (v55) at (9.5,2.5) {};
\node (v56) at (9.5,0) {};
\node (v58) at (13.5,0) {};
\node (v57) at (13.5,2.5) {};
\node (v60) at (-0.5,2.5) {};
\node (v59) at (-0.5,0) {};
\node (v61) at (5.5,2.5) {};
\node (v62) at (5.5,0) {};
\node (v63) at (11.5,0) {};
\node (v64) at (11.5,2.5) {};
\draw[fill=black] (-0.5,1.25) circle (2pt);
\draw[fill=black] (-0.1,1.25) circle (2pt);
\draw[fill=black] (-0.9,1.25) circle (2pt);
\draw[fill=black] (5.5,1.25) circle (2pt);
\draw[fill=black] (5.1,1.25) circle (2pt);
\draw[fill=black] (5.9,1.25) circle (2pt);
\draw[fill=black] (11.5,1.25) circle (2pt);
\draw[fill=black] (11.1,1.25) circle (2pt);
\draw[fill=black] (11.9,1.25) circle (2pt);
\end{tikzpicture}%
}

\usepackage[utf8]{inputenc}
\usepackage{graphicx}
\usepackage{dcolumn}
\usepackage{bm}

\usepackage{todonotes}

\begin{document}

\title{An Infinite Family of Elliptic Ladder Integrals}

\author[a,b]{Andrew McLeod,}
\emailAdd{a.mcleod@cern.ch}
\author[c]{Roger Morales,}
\emailAdd{roger.morales@nbi.ku.dk}
\author[c]{Matt von Hippel,}
\emailAdd{mvonhippel@nbi.ku.dk}
\author[c]{Matthias Wilhelm,}%
\emailAdd{matthias.wilhelm@nbi.ku.dk}
\author[c]{and Chi Zhang\begin{CJK*}{UTF8}{gbsn} (张驰)\end{CJK*}
}%
\emailAdd{chi.zhang@nbi.ku.dk}

\affiliation[a]{%
CERN, Theoretical Physics Department, 1211 Geneva 23, Switzerland}

\affiliation[b]{Mani L. Bhaumik Institute for Theoretical Physics, Department of Physics and Astronomy,
UCLA, Los Angeles, CA 90095, USA}

 \affiliation[c]{%
Niels Bohr International Academy, Niels Bohr Institute, Copenhagen University, Blegdamsvej 17, 2100 Copenhagen \O{}, Denmark}

\date{\today}%

\abstract{We identify two families of ten-point Feynman diagrams that generalize the elliptic double box, and show that they can be expressed in terms of the same class of elliptic multiple polylogarithms to all loop orders. Interestingly, one of these families can also be written as a $d$log form. For both families of diagrams, we provide new $2\ell$-fold integral representations that are linearly reducible in all but one variable and that make the above properties manifest. We illustrate the simplicity of this integral representation by directly integrating the three-loop representative of both families of diagrams. These families also satisfy a pair of second-order differential equations, making them ideal examples on which to develop bootstrap techniques involving elliptic symbol letters at high loop orders.}

\maketitle

\section{Introduction}

At large loop orders and high particle multiplicity, Feynman diagrams are generically expected to give rise to integrals over exceedingly complicated algebraic varieties. This behavior has already been observed in several classes of examples, such as the banana~\cite{SABRY1962401,Broadhurst:1993mw,Caffo:1998du,Laporta:2004rb,Muller-Stach:2011qkg,Adams:2013nia,Remiddi:2013joa,Bloch:2013tra,Adams:2014vja,Adams:2015gva,Adams:2015ydq,Remiddi:2016gno,Bloch:2016izu,Adams:2017ejb,Broedel:2017siw,Broedel:2018iwv,Bogner:2019lfa,Broedel:2019kmn,Klemm:2019dbm,Bonisch:2020qmm,Campert:2020yur,Frellesvig:2021vdl,Bonisch:2021yfw,Lairez:2022zkj,mirrors_and_sunsets}, traintrack~\cite{Bourjaily:2018ycu,Duhr:2022pch}, and tardigrade diagrams~\cite{Bourjaily:2018yfy}, all of which give rise to varieties of arbitrarily large dimension at high loop orders. The question of how to work with these types of integrals constitutes one of the major conceptual challenges currently faced in perturbative amplitude calculations. While important progress has been made towards developing the requisite analytic and numerical control over the banana integrals~\cite{Klemm:2019dbm,Bonisch:2020qmm,Bonisch:2021yfw,Lairez:2022zkj,Pogel:2022ken}, other examples---such as the two-loop tardigrade integral, which involves an integral over a K3 surface---have received considerably less attention (although recent progress has been made; see for instance~\cite{Lairez:2022zkj}).

The situation is significantly better for Feynman diagrams that only give rise to integrals over one-dimensional algebraic varieties, or in which no algebraic factors arise at all. Diagrams that fall into the latter class can be expressed in terms of multiple polylogarithms (MPLs)~\cite{Chen,G91b,Goncharov:1998kja,Remiddi:1999ew,Borwein:1999js,Moch:2001zr}, which enjoy a well-understood coaction structure that allows the analytic features of these functions to be systematically studied~\cite{Goncharov:2005sla,Goncharov:2010jf,Brown:2011ik,Brown1102.1312,Duhr:2011zq,Duhr:2012fh}, and for which efficient numerical evaluation algorithms exist~\cite{Bauer:2000cp,Vollinga:2004sn}. The simplest Feynman integrals that cannot be expressed in terms of MPLs are those that give rise to integrals over a single elliptic curve. These diagrams can generally be expressed in terms of elliptic multiple polylogarithms (eMPLs)~\cite{BeilinsonLevin,LevinRacinet2007,brown2011multiple,Broedel:2014vla,Broedel:2017kkb,Broedel:2017siw,Broedel:2018qkq}, which contain MPLs as special cases and share much of their mathematical structure. In particular, eMPLs are endowed with a coaction structure~\cite{2015arXiv151206410B,Broedel:2018iwv,Kristensson:2021ani,Forum:2022lpz,Wilhelm:2022wow}, and their numeric value can be efficiently computed for rational values of their arguments~\cite{Manin,2014arXiv1407.5167B,Broedel:2018iwv,Duhr:2019rrs,Weinzierl:2020fyx,Walden:2020odh}. 

While much of the desirable technology for working with eMPLs has recently been developed, only a handful of Feynman diagrams have in fact been evaluated in terms of these functions (see~\cite{Bourjaily:2022bwx} for a recent review). In part, this is because there is no general understanding of what kinds of functions any given Feynman diagram will evaluate to. A relatively primitive way to probe this question is by computing a diagram's leading singularity~\cite{CaronHuot:2012ab,Frellesvig:2017aai,Bourjaily:2018ycu,Bourjaily:2019hmc,Vergu:2020uur}; a Feynman integral can be expressed in terms of MPLs only if its leading singularity is at most algebraic (that is, if it does not involve any transcendental integrals). Similarly, Feynman integrals can only be expressed in terms of eMPLs if their leading singularity involves at most an integral over an elliptic curve (and not integrals over a higher-dimensional variety). However, in neither case does this criterion constitute a sufficient condition---see for instance the counterexamples in ref.~\cite{Duhr:2020gdd}. There also exist conditions that are sufficient but not necessary for an integral to be expressible in terms of MPLs, such as linear reducibility. Linear reducibility can be checked algorithmically via the polynomial reduction algorithm of refs.~\cite{Brown:2008um,Brown:2009ta,Panzer:2014gra}, but is a parametrization-dependent criterion~\cite{Panzer:2014gra,Heller:2019gkq,Duhr:2021fhk}. 

Being able to predict the types of functions that appear in specific classes of Feynman integrals is important from both a phenomenological and a theoretical point of view. Phenomenologically, it informs the type of technology that must be developed in order to efficiently evaluate these integrals numerically. On the theoretical side, our expectation as to what types of functions will appear has an impact on the methods used to evaluate these integrals analytically. For instance, our growing appreciation that large classes of amplitudes and diagrams are expressible in terms of MPLs, which includes examples to all loop orders and particle multiplicity (see for instance~\cite{Caron-Huot:2011zgw,Bourjaily:2018aeq,Caron-Huot:2018dsv,Bourjaily:2019iqr,Bourjaily:2019gqu,Caron-Huot:2020bkp,Bourjaily:2020qca,Bourjaily:2022tep}), has played a key role in the development of perturbative bootstrap methods, which have given rise to some of the highest-loop amplitude results in the literature~\cite{Dixon:2011pw,Dixon:2011nj,Dixon:2013eka,Dixon:2014iba,Dixon:2014voa,Dixon:2015iva,Dixon:2016apl,Li:2016ctv,Caron-Huot:2016owq,Dixon:2016nkn,Chicherin:2017dob,Almelid:2017qju,Drummond:2018caf,Caron-Huot:2019vjl,Dixon:2020cnr,Golden:2021ggj,Brandhuber:2012vm,Dixon:2020bbt,Dixon:2022rse,Dixon:2022xqh}. 
Conversely, only a small number of Feynman diagrams and amplitudes are currently known to be expressible in terms of eMPLs.

In this paper, we identify the first infinite family of Feynman diagrams that can be expressed in terms of the same class of eMPLs to all loop orders. These diagrams generalize the ten-point double-box integral through the addition of loops both to the left and the right of the massless pair of external legs, as depicted in Figure~\ref{fig:LRdiagram_p}. While the double box integral has long been known to be elliptic~\cite{Paulos:2012nu,CaronHuot:2012ab,Nandan:2013ip}, we here show that the same elliptic curve appears in the leading singularity of this family of integrals at all loop orders. Moreover, we show that these diagrams can be expressed as $2\ell$-fold integrals that are linearly reducible in all but one integration variable, where only a single square root---the elliptic curve---obstructs linear reducibility in the last variable.\footnote{This is also known in the literature as a \emph{linearly reducible elliptic Feynman integral}, see ref.~\cite{Hidding:2017jkk}.} This implies that these integrals can all be evaluated in terms of elliptic polylogarithms that involve the same elliptic curve. 

We also consider a second family of Feynman diagrams, in which the pair of massless legs are attached to different rungs of the ladder, as depicted in Figure~\ref{fig:LMRdiagram_p}.  The leading singularity of these diagrams does not involve an elliptic curve; however, it can be seen at low loop orders that they involve the same elliptic curve as the previous family of diagrams. We also provide $2\ell$-fold integral expressions for this family of integrals that are linearly reducible in all but one integration variable. 

The Feynman-parametrized integral representations that we provide for both families of diagrams are ideally suited to direct integration. We illustrate this fact by computing the three-loop representative of each family, in the first case in terms of eMPLs, and in the second case as a one-fold integral over MPLs. The fact that these integrals can be readily evaluated at low loop orders makes it possible to study the properties of their symbol alphabets, and in particular whether these alphabets saturate at some fixed loop order. If such a saturation were observed, it would make it likely that these integrals could be bootstrapped to much higher loop orders, or even resummed in the coupling, as has been done for other families of ladder integrals~\cite{Broadhurst:2010ds,Caron-Huot:2018dsv,He:2020uxy}. This possibility is further enhanced by the existence of a pair of on-shell differential equations that relate these integrals at adjacent loop orders~\cite{Drummond:2010cz}. In the present paper, we derive the explicit form of these differential equations, but leave the further tantalizing possibility of bootstrapping these integrals to future work.

The paper is organized as follows. In section~\ref{sec:review}, we review MPLs and the notion of linear reducibility, which will play an important role in this work. Section~\ref{sec:ell_ladder} introduces the first family of integrals we study, and here we demonstrate how to render these integrals linearly reducible (in all but one parameter) and how to integrate them into eMPLs. We also derive the pair of differential equations that they satisfy. In section~\ref{sec:staggered}, we carry out an analogous analysis of our second family of integrals. We finally conclude and discuss further directions in section~\ref{sec:conclusion}. 

We also include two appendices in this work. Appendix~\ref{app: LS} shows that removing a rung from a ladder diagram does not change its leading singularity, as long as no external legs are attached to this rung. We use this result to conclude that the leading singularities of the families of diagrams we study are the same at all loop orders. Appendix~\ref{app: DEs} showcases the differential equations these integrals satisfy, and in particular shows that they are satisfied at the level of the integrand for the Feynman parameter integral representation we present for the first family of diagrams. We additionally include ancillary files that contain the pair of three-loop examples we study in sections~\ref{sec:ell_ladder} and~\ref{sec:staggered}, expressed as a one-fold integral over MPLs and in terms of eMPLs. These files can be accessed at~\cite{ancillary_file}.

While this work was in progress, we were made aware of related upcoming work \cite{Cao:2023tpx}, which considers similar families of elliptic ladder integrals.

\section{Multiple Polylogarithms and Linear Reducibility}
\label{sec:review}

The simplest Feynman integrals can be evaluated in terms of MPLs, which correspond to iterated integrals over certain types of $d$log forms. One of the standard notations used for these functions is given by the recursive definition
\begin{equation}
    G(a_{1},\ldots,a_{n};z):=\int_{0}^{z}d\log(t-a_{1})\, G(a_{2},\ldots,a_{n};t)\,, \label{Gpolylot} 
\end{equation} 
where $G(;z):=1$ and the arguments $a_i$ and $z$ are allowed to be algebraic. When $a_1 = \cdots = a_n = 0$ these functions must be regulated, which is done by defining
\begin{equation}
    G(\underbrace{0, \ldots, 0}_{n} ; z):=\frac{1}{n !} \, (\log z)^{n}
\end{equation} 
as a special case. The number of indices $n$ is referred to as the weight of each MPL.

The space of MPLs respects a Hopf algebra structure~\cite{Goncharov:2005sla}, which allows us to associate an object called a \textit{symbol} to each function~\cite{Goncharov:2010jf, Duhr:2011zq}. The symbol of a weight-$n$ function $F^{(n)}$ whose total differential is
\begin{equation} \label{Gdiff}
    d F^{(n)}=\sum_i F^{(n-1)}_i \, d\log s_i \, 
\end{equation}
is recursively defined by
\begin{equation} 
\mathcal{S}(F^{(n)}):=\sum_i \mathcal{S}(F^{(n-1)}_i)\otimes s_i\,,
\end{equation} 
 where the recursion terminates with $\mathcal{S}(1) = 1$. The logarithmic arguments $s_{i}$ that appear in this recursion are called \emph{symbol letters}, while the total collection of symbol letters that appear in a function's symbol is referred to as that function's \emph{symbol alphabet}. For more details on the Hopf algebra structure of MPLs and symbols, see for instance~\cite{Duhr:2014woa}.

In general, it is hard to determine whether an integral can be evaluated in terms of MPLs without explicitly carrying out each of the integrations. However, in special cases one can prove that an integral is \emph{linearly reducible}, which guarantees it can be integrated into MPLs. Linear reducibility can be checked using a polynomial 
reduction algorithm~\cite{Brown:2008um,Brown:2009ta,Panzer:2014gra}, which proceed as follows.\footnote{This condition can also be checked graphically, using compatibility graph reduction~\cite{Brown:2009ta}.} Suppose one has an integral of the form
\begin{equation}
\int_{[0,\infty]^{n}} \frac{d^{n} \vec{\alpha}}{\prod_{i=1}^{m}f_{i}(\vec{\alpha}, \vec{x})} \,,   
\end{equation}
where each of the factors $f_{i}(\vec{\alpha}, \vec{x})$ are polynomials in the integration variables $\vec{\alpha}$ and some set of auxiliary variables $\vec{x}$, with rational coefficients. One begins by identifying one of the integration variables $\alpha_e$ that appears at most linearly in each of the polynomials $f_i$. This implies that each of the factors in this integral can be written as $f_{i}= A_{i} \, \alpha_{e}+B_{i}$. Then, one can mimic carrying out the integral over $\alpha_{e}$ by defining a new set of polynomials
\begin{equation} 
\{A_{k}, B_{k}, A_{i}B_{j}-A_{j}B_{i}  \,  \vert  \,  1\leq k \leq m, 1\leq i<j\leq m  \} \setminus \{0\} \,,
\end{equation} 
which no longer depend on $\alpha_e$. One then repeats this procedure by looking for another integration variable that appears at most linearly in the set of irreducible factors of these new polynomials, which can be eliminated in the same way. If there exists some sequence in which all integration variables can be eliminated in this way, the original integral is linearly reducible, and can be evaluated in terms of MPLs by carrying out the integrations in the same order. Moreover, the symbol alphabet of the result is contained within the final set of polynomials generated by this procedure.

\section{The Elliptic Ladder Integrals}
\label{sec:ell_ladder}

\begin{figure}
\centering
\begin{tikzpicture}[scale=0.68, label distance=-1mm]
\node (v20) at (-3.5,2.5) {};
\node (v26) at (-1.5,2.5) {};
\node (v27) at (0.5,2.5) {};
\node at (2.5,2.5) {};
\node (v35) at (4.5,2.5) {};
\node (v39) at (6.5,2.5) {};
\node (v42) at (8.5,2.5) {};
\node (v23) at (-3.5,0) {};
\node (v25) at (-1.5,0) {};
\node (v28) at (0.5,0) {};
\node (v38) at (2.5,0) {};
\node (v36) at (4.5,0) {};
\node (v40) at (6.5,0) {};
\node (v41) at (8.5,0) {};
\node (v19) at (-5,3) {};
\node (v21) at (-4,4) {};
\node (v22) at (-5,-0.5) {};
\node (v24) at (-4,-1.5) {};
\node (v45) at (10,-0.5) {};
\node (v44) at (10,3) {};
\node (v43) at (9,4) {};
\node (v46) at (9,-1.5) {};
\node (v37) at (2.5,4) {};
\node[label=left:\textcolor{blue!50}{$x_8$}] (v11) at (-4.5,1.25) {};
\node (v1) at (-2.5,1.25) {};
\node (v12) at (-1.5,1.25) {};
\node (v13) at (0.5,1.25) {};
\node (v4) at (1.5,1.25) {};
\node (v9) at (3.5,1.25) {};
\node (v16) at (4.5,1.25) {};
\node (v17) at (6.5,1.25) {};
\node (v7) at (7.5,1.25) {};
\node[label=right:\textcolor{blue!50}{$x_3$}] (v18) at (9.5,1.25) {};
\node[label=above:\textcolor{blue!50}{$x_{10}$}] (v2) at (-0.5,4) {};
\node[label=below:\textcolor{blue!50}{$x_6$}] (v3) at (-0.5,-1.5) {};
\node[label=above:\textcolor{blue!50}{$x_1$}] (v8) at (5.5,4) {};
\node (v47) at (-2.5,2.5) {};
\node (v48) at (-2.5,0) {};
\node (v52) at (1.5,0) {};
\node (v49) at (1.5,2.5) {};
\node (v55) at (3.5,2.5) {};
\node (v56) at (3.5,0) {};
\node (v58) at (7.5,0) {};
\node (v57) at (7.5,2.5) {};
\node (v60) at (-0.5,2.5) {};
\node (v59) at (-0.5,0) {};
\node (v63) at (5.5,0) {};
\node (v64) at (5.5,2.5) {};
\node at (-0.5,1.25) {};
\node at (2.5,1.25) {};
\node at (5.5,1.25) {};
\node (v5) at (2.5,-1.5) {};
\node[label=below:\textcolor{blue!50}{$x_5$}] (v6) at (5.5,-1.5) {};
\draw[very thick]  (v19.center) edge (v20.center);
\draw[very thick]  (v20.center) edge (v21.center);
\draw[very thick]  (v20.center) edge (v23.center);
\draw[very thick]  (v23.center) edge (v22.center);
\draw[very thick]  (v23.center) edge (v24.center);
\draw[very thick]  (v23.center) edge (v25.center);
\draw[very thick]  (v20.center) edge (v26.center);
\draw[very thick]  (v26.center) edge (v25.center);
\draw[very thick]  (v28.center) edge (v27.center);
\draw[very thick]  (v27.center) edge (v35.center);
\draw[very thick]  (v35.center) edge (v36.center);
\draw[very thick]  (v36.center) edge (v28.center);
\draw[very thick]  (v38.center) edge (v37.center);
\draw[very thick]  (v38.center) edge (v5.center);
\draw[very thick]  (v39.center) edge (v40.center);
\draw[very thick]  (v40.center) edge (v41.center);
\draw[very thick]  (v41.center) edge (v42.center);
\draw[very thick]  (v39.center) edge (v42.center);
\draw[very thick]  (v42.center) edge (v43.center);
\draw[very thick]  (v42.center) edge (v44.center);
\draw[very thick]  (v41.center) edge (v45.center);
\draw[very thick]  (v41.center) edge (v46.center);
\draw[densely dashed,very thick]  (v26.center) edge (v27.center);
\draw[densely dashed,very thick]  (v25.center) edge (v28.center);
\draw[densely dashed,very thick]  (v35.center) edge (v39.center);
\draw[densely dashed,very thick]  (v36.center) edge (v40.center);
\draw[thick,blue!50]  (v11.center) edge (v12.center);
\draw[thick,blue!50]  (v13.center) edge (v16.center);
\draw[thick,blue!50]  (v17.center) edge (v18.center);
\draw[thick,blue!50]  (v1.center) edge (v48.center);
\draw[thick,blue!50]  (v48.center) edge (v3.center);
\draw[thick,blue!50]  (v3.center) edge (v52.center);
\draw[thick,blue!50]  (v52.center) edge (v4.center);
\draw[thick,blue!50]  (v1.center) edge (v47.center);
\draw[thick,blue!50]  (v47.center) edge (v2.center);
\draw[thick,blue!50]  (v2.center) edge (v49.center);
\draw[thick,blue!50]  (v49.center) edge (v4.center);
\draw[thick,blue!50]  (v55.center) edge (v56.center);
\draw[thick,blue!50]  (v56.center) edge (v6.center);
\draw[thick,blue!50]  (v6.center) edge (v58.center);
\draw[thick,blue!50]  (v58.center) edge (v57.center);
\draw[thick,blue!50]  (v57.center) edge (v8.center);
\draw[thick,blue!50]  (v8.center) edge (v55.center);
\draw[densely dashed,thick,blue!50]  (v12.center) edge (v13.center);
\draw[densely dashed,thick,blue!50]  (v2.center) edge (v3.center);
\draw[densely dashed,thick,blue!50]  (v16.center) edge (v17.center);
\draw[densely dashed,thick,blue!50]  (v8.center) edge (v6.center);
\draw[fill=blue!50] (v1) circle (2pt);
\draw[fill=blue!50] (-0.5,1.25) circle (2pt);
\draw[fill=blue!50]  (v4) circle (2pt);
\draw[fill=blue!50] (5.5,1.25) circle (2pt);
\draw[fill=blue!50]  (v9) circle (2pt);
\draw[fill=blue!50]  (v7) circle (2pt);
\node at (-3.25,3.5) {$p_9$};
\node at (-4.5,2.35) {$p_8$};
\node at (-4.5,0.15) {$p_7$};
\node at (-3.25,-1) {$p_6$};
\node at (2,-0.9) {$p_5$};
\node at (3.1,3.3) {$p_{10}$};
\node at (8.25,3.5) {$p_1$};
\node at (8.25,-1) {$p_4$};
\node at (9.5,0.15) {$p_3$};
\node at (9.5,2.35) {$p_2$};
\draw [thick,decoration={brace,mirror,raise=0.5cm},decorate] (-3.4,-2) -- (2.4,-2);
\node at (-0.5,-3.25) {$L$};
\draw [thick,decoration={brace,mirror,raise=0.5cm},decorate] (2.6,-2) -- (8.4,-2);
\node at (5.5,-3.25) {$R$};
\end{tikzpicture} 
\caption{The ten-point elliptic ladder diagrams, which are allowed to involve any number of loops on either side ($L$ and $R$) of the pair of massless legs associated with $p_5$ and $p_{10}$. The momentum-space diagram is shown in black, while the dual graph is shown in blue.}
\label{fig:LRdiagram_p}
\end{figure}
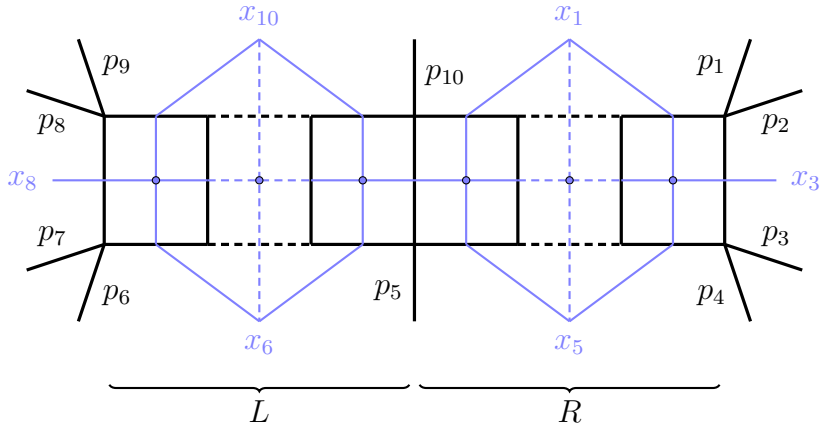

We begin by studying the family of Feynman diagrams depicted in Figure~\ref{fig:LRdiagram_p}, where the dashed lines indicate that any number of loops is allowed to appear on either side of the diagram. Each diagram that falls into this family can be identified by a pair of positive integers $L$ and $R$, which indicate the number of loops to the left and right of the massless external edges. 
Due to the symmetry that swaps $L$ and $R$, we restrict our attention to $L \ge R$. After taking into account this symmetry, there are $\lfloor \ell/2 \rfloor$ distinct diagrams in this family at $\ell$ loops. Anticipating the results we derive below, we refer to these diagrams as \emph{elliptic ladder diagrams}.

The elliptic ladder diagrams depend on the same set of kinematic variables for all $L$ and $R$. In particular, the numerator of these diagrams can be chosen such that the corresponding integrals respect a dual conformal symmetry~\cite{Bern:2006ew,Drummond:2006rz,Drummond:2007aua,Bern:2007ct,Alday:2007hr,Nguyen:2007ya,Drummond:2008vq}. This allows us to capture their kinematic dependence using dual coordinates, defined in terms of external momenta via the relation
\begin{equation}
x_i - x_{i+1} = p_i \, .
\end{equation}
We furthermore denote the squared differences between dual coordinates by
\begin{equation}
x_{i,j}^2 = (x_i - x_j)^2 = (p_i + \dots + p_{j-1})^2 \, .
\end{equation}
In terms of these variables, the elliptic ladder integrals read
\begin{equation}  \label{ell_ladder_integrand}
\mathcal{I}_{L,R}=\int \frac{d^4 x_{l_1} \dots d^4 x_{l_{R+L}} \quad x_{3,8}^{2} \, (x_{1,5}^{2})^R \, (x_{6,10}^{2})^L}{x_{l_1,3}^{2} \Big[ \prod_{i=1}^R x_{l_i, 1}^2 \, x_{l_i, 5}^2 \, x_{l_i, l_{i+1}}^2 \Big] \, \Big[ \prod_{i=R+1}^{R+L} x_{l_i, 10}^2 \, x_{l_i, 6}^2 \, x_{l_i, l_{i+1}}^2 \Big] x_{l_{R+L},8}^{2}}\,,
\end{equation}
where $x_{l_i}$ is the dual point associated to the $i$-th loop momentum, and we have introduced $x_{l_{R+L},l_{R+L+1}}^2=1$ for notational compactness. Note that the numerator differs by an overall factor from the one used in ref.\ \cite{Kristensson:2021ani} for the elliptic double-box integral, which was chosen due to its relation with the one-loop hexagon integral in six dimensions.

The kinematic dependence of the elliptic ladder diagrams can be expressed in terms of just seven dual-conformal-invariant cross-ratios to all loop orders:
\begin{gather}
u_1=\frac{x_{1,3}^2 \, x_{5,8}^2}{x_{1,5}^2 \, x_{3,8}^2} \, , \quad u_2=\frac{x_{3,6}^2 \, x_{8,10}^2}{x_{3,8}^2 \, x_{6,10}^2} \, , \quad u_3 =  \frac{x_{1,3}^2 \, x_{5,10}^2}{x_{1,5}^2 \, x_{3,10}^2} \, , \quad u_4=\frac{x_{1,6}^2 \, x_{3,5}^2}{x_{1,5}^2 \, x_{3,6}^2} \, ,\\ 
u_5=\frac{x_{1,5}^2 \, x_{6,10}^2}{x_{1,6}^2 \, x_{5,10}^2} \, , \quad  v_1=\frac{x_{1,8}^2 \, x_{3,5}^2}{x_{1,5}^2 \, x_{3,8}^2} \, , \quad v_2=\frac{x_{3,10}^2 \, x_{6,8}^2}{x_{3,8}^2 \, x_{6,10}^2} \, .
\end{gather}
In addition, the integral $\mathcal{I}_{L,R}$ depicted in Figure~\ref{fig:LRdiagram_p} has two reflection symmetries. We denote the action of reflection across the vertical axis by $\mathcal{V}$, and the action of reflection across the horizontal axis by $\mathcal{H}$. That is, we have $\mathcal{V}(\mathcal{I}_{L,R})=\mathcal{I}_{R,L}$ and $\mathcal{H}(\mathcal{I}_{L,R})=\mathcal{I}_{L,R}$, where these reflections act on the cross-ratios as follows:
\begin{align}
 &\mathcal{V}:\quad u_1 \leftrightarrow u_2,\,v_1 \leftrightarrow v_2,\, u_3 \rightarrow u_2 u_4/v_1,\, u_4 \rightarrow u_3 v_2 / u_1\,,
 \label{eq: vertical symmetry}\\
 &\mathcal{H}:\quad u_1 \leftrightarrow v_1,\, u_2 \leftrightarrow v_2, \,u_3 \leftrightarrow u_4\, .
 \label{eq: horizontal symmetry}
\end{align}
The variable $u_5$ is left invariant by both symmetry transformations.

\subsection{Feynman Parametrization and Linear Reducibility}
\label{subsec: Feynman ladders}

While the two-loop elliptic ladder diagram has already received a large amount of attention over the last decade~\cite{Paulos:2012nu,CaronHuot:2012ab,Nandan:2013ip,Bourjaily:2017bsb,Kristensson:2021ani,Wilhelm:2022wow,Morales:2022csr}, diagrams in this family involving three or more loops have not been previously studied. However, they correspond to multi-soft limits of the traintrack diagrams considered in ref.\ \cite{Bourjaily:2018ycu}, for which a $2\ell$-fold integral representation was derived using loop-by-loop Feynman parametrization. We can thus derive a  $2\ell$-fold integral representation of the elliptic ladder diagrams by taking the appropriate limit of the result presented there, which just involves identifying the appropriate external dual points. Doing so, we find
\begin{equation}
\mathcal{I}_{L,R} = \int_0^\infty d^\ell \vec{\alpha} \ d^\ell \vec{\beta} \frac{1}{f_1 \dots f_R \,  f_{R+1} \dots f_{R+L} \, g_\ell} \, ,
\label{eq:lblexpression}
\end{equation}
where
\begin{align}
\hspace{-0.2cm} f_{k \leq R} &= f_{k-1} + \alpha_k + \beta_k + \alpha_k \beta_k + \sum_{j=1}^{k-1} \Big( \alpha_j \beta_k + \alpha_k \beta_j \Big), \\
\hspace{-0.2cm} f_{R+1} &= u_3 u_4 u_5 f_R + \alpha_{R+1} + \beta_{R+1} + \alpha_{R+1} \beta_{R+1} + \sum_{j=1}^{R} \Big( u_4 \alpha_j \beta_{R+1} + u_3 \alpha_{R+1} \beta_j \Big),
\end{align}
\vspace{-0.5cm}
\begin{align}
&\hspace*{-0.2cm} f_{k > R+1} = f_{k-1} + \alpha_k + \beta_k + \alpha_k \beta_k + \sum_{j=1}^{R} \Big( u_4 \alpha_j \beta_k + u_3 \alpha_k \beta_j \Big) + \sum_{j=R+1}^{k-1} \Big( \alpha_j \beta_k + \alpha_k \beta_j \Big), \label{eq:bigfdef}\\
&\hspace*{-0.2cm} g_\ell = 1 + \sum_{j=1}^R \Big( v_1 \alpha_j + u_1 \beta_j \Big) + \sum_{j=R+1}^\ell \Big( u_2 \alpha_j + v_2 \beta_j \Big),
\end{align}
with $f_0=0$. 

While this Feynman parametrization is not linearly reducible, it provides a starting point for finding a linearly reducible parametrization via changes of variables.
Notably, there are no cross terms of the form $\beta_k \beta_j$ in the polynomials that appear in this integral representation, which means we can carry out linear changes of variables within the space of $\beta_j$ variables without generating quadratic or higher powers in $\beta_k$. Let us therefore apply the linear transformation
\begin{subequations} \label{eq:change_variables_LR}
\begin{align}
\beta_{1< k \leq R} &\to \beta_k - \beta_{k-1}\, ,\\
\beta_{R+1} &\to \beta_{R+1} - \frac{u_1}{v_2} \beta_R - \beta_{R+2}\, , \\
\beta_{R+1< k < \ell} &\to \beta_k - \beta_{k+1}\, .
\end{align}
\end{subequations}
After this change of variables, we find that $\mathcal{I}_{L,R}$ is linearly reducible in all but one variable, with just a single square root of a quartic polynomial in the last integration variable appearing in the last step.\footnote{We have explicitly checked this linear reducibility using the {\Maple} package {\HyperInt}~\cite{Panzer:2014caa} up to six loops, but conjecture it holds to all loop orders. It would be interesting to find a general proof for this property.} We can thus evaluate the first $2\ell-1$ integrals in terms of MPLs, using the integration order $\{ \alpha_1,\dots,\alpha_\ell,\beta_\ell,\dots,\beta_{R+2},\beta_1,\dots,\beta_{R+1} \}$. 

We denote the final integration variable $\beta_{R+1}$ as $x$, and find that the square root $y$ at any loop order is defined via
\begin{align}
\label{eq: elliptic curve}
y^2 
&= \bigg(\frac{1}{v_2^2} \big[ (1-v_1+v_2 x) (v_2 u_3 - u_1)+u_1 u_2 (1-u_4) \big] +h_1+h_2 \bigg)^2 - 4 h_1 h_2\, ,
\end{align}
where
\begin{align}
h_1 &= \frac{1}{v_2^2} \big[ (v_2 x+ u_1) (u_2 u_4 - v_1)+v_1 v_2 (u_3 - 1) \big]\, ,\\
h_2 &= \frac{-1}{v_2} \big[ v_2 x^2 + (1-u_2+v_2) x + 1 \big].
\end{align}
By computing the $j$-invariant~\cite{Broedel:2017kkb} of this curve, namely
\begin{equation}
j=256 \,\frac{(1-\lambda (1- \lambda) )^3}{\lambda^2 (1-\lambda)^2} \, ,
\end{equation}
where
\begin{equation}
\lambda= \frac{r_{14} \, r_{23}}{r_{13} \, r_{24}} \, , \qquad r_{ij}=r_i - r_j\, ,
\end{equation}
and $r_i$ are the roots of the elliptic curve, it can be seen that this curve is isomorphic to the one found in ref.\ \cite{Kristensson:2021ani} for the elliptic double-box diagram.

Based on the two-loop elliptic double box, as well as the fact that the same leading singularity appears for the elliptic ladder diagrams for any $L$ and $R$ (which we prove in appendix~\ref{app: LS}), we expect $\mathcal{I}_{L,R}$ to take the form
\begin{equation} \label{eq: ell_ladder_one_fold}
\mathcal{I}_{L,R} = \int_0^\infty \frac{dx}{y} \, \mathcal{G}_{L,R}^{(2\ell-1)}(x,y)\, ,
\end{equation}
where $\mathcal{G}_{L,R}^{(2\ell-1)}(x,y)$ is a linear combination of MPLs of uniform weight $2\ell-1$ involving only rational numerical coefficients. In particular, since the leading singularity of every one of these integrals is given by the exact same expression, the elliptic curve $y$ that appears in eq.~\eqref{eq: ell_ladder_one_fold} will be the same for all $L$ and $R$.

\subsection{Integration to Elliptic Multiple Polylogarithms}
\label{subsec:ell_integration}

Due to the appearance of the elliptic curve in the final integral, we must work within the larger class of eMPLs. One formalism for this space of functions is defined recursively by the equation~\cite{Broedel:2017kkb}
\begin{equation}
\label{Eiterateddefinition}
    \Ef{n_1 & \ldots & n_k}{c_1 & \ldots& c_k}{x}=
    \int_{0}^{x}d x'\,\psi_{n_{1}}(c_{1},x') \, \Ef{n_{2} & \ldots & n_k}{c_{2} & \ldots& c_k}{x'} \, ,
\end{equation}
where $\mathrm{E}_{4}(;x)=1$. Similarly to eq.~\eqref{Gpolylot}, this definition diverges when $c_k = 0$, but these divergences can be regulated in an analogous way to MPLs, as shown in ref.~\cite{Broedel:2017kkb}. However, unlike the case of MPLs, an infinite tower of integration kernels with simple poles $\psi_{n}$ must be defined to span the space of functions in $x$ and $y$ that can be generated in the integrand by partial fractioning and integration-by-parts identities. Luckily, as we will see below, the elliptic ladder integrals can be integrated using just the kernels
\begin{subequations} \label{psibasis}
    \begin{align}
        &\psi_{0}(0,x)=\frac{1}{y} \,,\qquad  \:\:\psi_{-1}(\infty,x)=\frac{x}{y}\,, \\
        &\psi_{1}(c,x)=\frac{1}{x-c}\,,\quad 
        \psi_{-1}(c,x)=\frac{y_{c}}{y(x-c)}\,,
    \end{align} 
\end{subequations}
where $y_{c}=y\vert_{x=c}$.

In order to compute $\mathcal{I}_{L,R}$ in terms of $\rm{E}_{4}$ functions, we must rewrite each polylogarithm in $\mathcal{G}^{(2\ell-1)}(x,y)$ in terms of eMPLs. Using the differential property of MPLs from eq.~\eqref{Gdiff}, we have that
\begin{equation}
    d\mathcal{G}^{(2\ell-1)}(x,y)=\sum_{i} \mathcal{G}^{(2\ell-2)}_i(x,y) \, d\log R_{i}(x,y)\,, 
\end{equation} 
for some set of lower-weight functions $\mathcal{G}^{(2\ell-2)}_i(x,y)$, where $R_{i}(x,y)$ are rational functions of $x$ and $y$. This allows us to rewrite $\mathcal{G}^{(2\ell-1)}(x,y)$ as
\begin{equation} \label{fibra_expansion}
    \mathcal{G}^{(2\ell-1)}(x,y)= \mathcal{G}^{(2\ell-1)}(0,y_{0})+\sum_{i} \int_{0}^{x}  \mathcal{G}^{(2\ell-2)}_i(t,y_{t}) \, \partial_{t} \log R_{i}(t,y_{t})  \, dt \, ,
\end{equation} 
and then to re-express each of the $d$logs $\partial_{t} \log R_{i}(t,y_{t})$ as a linear combination of the kernels $\psi_{0,\pm1}$. This can always be done, since these $d$logs have only simply poles in $t$, and are rational in $t$ and $y_{t}$. Applying this type of expansion recursively to the functions $\mathcal{G}^{(2\ell-2)}_i(x,y)$ allows us to convert $\mathcal{G}^{(2\ell-1)}(x,y)$ into an equivalent expression in the $\rm E_{4}$ notation. Once we have done that, the last integration in eq.~\eqref{eq: ell_ladder_one_fold} is trivial, since $\int dx/y$ is simply $\int \psi_{0}\,dx$.

While the evaluation of the elliptic ladder integrals in terms $\rm E_{4}$ functions can be straightforwardly carried out, the resulting expression hides the fact that they are pure functions---namely, that they can be written in such a way that the kinematic dependence only appears in the eMPLs, and not in the prefactors multiplying these functions. One way to make this purity manifest is to convert the eMPL result into $\widetilde{\Gamma}$ functions~\cite{Broedel:2017kkb,Broedel:2018iwv}, which are iterated integrals on a torus with modular parameter $\tau$, defined by
\begin{equation}
    \gamt{n_1 & \ldots & n_k}{w_1 & \ldots& w_k}{w|\tau}=
    \int_{0}^{w} d w'\,g^{(n_{1})}(w'{-}w_{1},\tau) \, \gamt{n_{2} & \ldots & n_k}{w_{2} & \ldots& w_k}{w'|\tau} \, ,
\end{equation}
with $\widetilde{\Gamma}(;w|\tau)=1$. Iterated integrals of this type are said to have length $k$ and weight $\sum_k n_k$. 
The integration kernels $g^{(n)}(w,\tau)$ are generated by the \emph{Eisenstein-Kronecker series}
\begin{equation} \label{eq:EK_series}
    \frac{\partial_{w}\theta_{1}(0|\tau) \, \theta_{1}(w+\alpha|\tau)}{\theta_{1}(w|\tau) \, \theta_{1}(\alpha|\tau)} = \sum_{n\geq 0}\alpha^{n-1}g^{(n)}(w,\tau)\, ,
\end{equation}
where $\theta_{1}(w|\tau)$ is the odd Jacobi theta function.

The translation between $\rm E_{4}$ functions and $\widetilde{\Gamma}$ functions depends on the bijection used between the elliptic curve and the torus. Here we adopt the conventions of refs.~\cite{Kristensson:2021ani,Wilhelm:2022wow}, and refer the reader there for further details.
We have performed this procedure explicitly for the three-loop case of $\mathcal{I}_{2,1}$ and packaged it into a more compact notation,
the so-called $\mathcal{E}_{4}$ functions \cite{Broedel:2018qkq}. They are defined in complete analogy to eq.~\eqref{Eiterateddefinition}:
\begin{equation}
\label{curlyEiterateddefinition}
    \cEf{n_1 & \ldots & n_k}{c_1 & \ldots& c_k}{x}=
    \int_{0}^{x}d  x'\,\Psi_{n_{1}}(c_{1},x') \, \cEf{n_{2} & \ldots & n_k}{c_{2} & \ldots& c_k}{x'} \, ,
\end{equation}
with $\mathcal{E}_{4}(;x)=1$ and
\begin{align}
 \Psi_{\pm (n>0)}(c,x) \, d  x=\Bigl(&g^{(n)}(w-w_{c}^{+})\pm g^{(n)}(w-w_{c}^{-})  
\nonumber \\
    &
    - \delta_{\pm n,1}\bigl( g^{(1)}(w-w_{\infty}^{+})+g^{(1)}(w-w_{\infty}^{-} )\bigr) \Bigr) d w \, .
\label{Psikernels} 
\end{align}
We also have $\Psi_0(x) \, d x=d w$, with $w_{c}^{\pm}$ denoting the torus images of $(c,\pm y_{c})$. 
We provide the resulting expression in an ancillary file, accessible at \cite{ancillary_file}.

\subsection{Differential Equations}
\label{subsec:diffeqs}

In ref.~\cite{Drummond:2010cz}, it was shown that many ladder-like integrals satisfy on-shell differential equations that relate diagrams at different loop orders. These differential equations rely on the fact that one of the dual points appears in just a single propagator. Applying the Laplacian operator with respect to that coordinate and pulling it under the integration sign, one obtains the delta function  
\begin{equation}
\Box_i \frac{1}{x_{ir}^2}=-4 i \pi^2 \delta^{(4)}(x_i-x_j)\,,
\end{equation}
which has the effect of trivializing one of the loop momentum integrations. One thereby obtains a second-order differential equation that relates diagrams at different loop orders.

With this in mind, let us apply a Laplacian operator with respect to the dual point $x_8$ to the integral $\mathcal{I}_{L,R}$. Expressing this differential equation in terms of dual-conformal cross-ratios, we find
\begin{equation}
\mathcal{D}_L \,  \mathcal{I}_{L,R}= \mathcal{I}_{L-1,R}\,,
\label{eq: differential equation left}
\end{equation}
where
\begin{align}
\mathcal{D}_L = u_2 v_2\Big[ &
-2 u_3 u_4 u_5  \left( \partial_{u_1}+\partial_{v_1}\right) - 2 \left(\partial_{u_2} +  \partial_{v_2} \right) \nonumber\\
 &- 
  u_1 u_3 u_4 u_5 \partial^2_{u_1} - 
  v_1 u_3 u_4 u_5 \partial^2_{v_1} - 
  u_2  \partial^2_{u_2} - 
  v_2 \partial^2_{v_2} \nonumber\\
 &+ 
  u_3 u_4 u_5 (1 - u_1 - v_1)  \partial_{v_1}\partial_{u_1} + 
  (1 - u_2 - v_2) \partial_{v_2}\partial_{u_2} \nonumber\\
 &- 
  (u_1 - u_3 + u_2 u_3 u_4 u_5) \partial_{u_2}\partial_{u_1} - 
  (v_1-u_4  + v_2 u_3 u_4 u_5 ) \partial_{v_2}\partial_{v_1} \nonumber\\
 &-   (u_1 + v_2 u_3 u_4 u_5 ) \partial_{v_2}\partial_{u_1}  - 
   (v_1 + u_2 u_3 u_4 u_5 )  \partial_{v_1}\partial_{u_2}     
\Big] \,.
\label{eq: differential operator}
\end{align}
Quite conveniently, it turns out that this differential equation can be checked at the level of the integrand, as we show in appendix \ref{app: DEs}.
Similarly, applying a Laplacian  with respect to $x_3$, we find \begin{equation}
\mathcal{D}_R \,  \mathcal{I}_{L,R}= \mathcal{I}_{L,R-1}\,,
\label{eq: differential equation right}
\end{equation}
where $\mathcal{D}_R$ can be obtained from eq.~\eqref{eq: differential operator} by symmetry, namely $\mathcal{D}_R=\mathcal{V}(\mathcal{D}_L)$. 
The boundary values for these differential equations, $\mathcal{I}_{0,R}$ and $\mathcal{I}_{L,0}$, are the well-known Usyukina-Davydychev ladders \cite{Usyukina:1993ch}. 

The analogous differential equation that is satisfied by the double-pentagon ladders has been solved for finite values of the coupling~\cite{Caron-Huot:2018dsv}. 
It would be interesting to investigate whether a similar approach is also possible here, in particular given the recent advances made in applying bootstrap methods to elliptic Feynman integrals~\cite{Morales:2022csr}.

\section{The Staggered Elliptic Ladder Integrals}
\label{sec:staggered}

\begin{figure}
\centering
\begin{tikzpicture}[scale=0.68, label distance=-1mm]
\node (v20) at (-3.5,2.5) {};
\node (v26) at (-1.5,2.5) {};
\node (v27) at (0.5,2.5) {};
\node (v31) at (2.5,2.5) {};
\node (v29) at (4.5,2.5) {};
\node (v33) at (6.5,2.5) {};
\node at (8.5,2.5) {};
\node (v35) at (10.5,2.5) {};
\node (v39) at (12.5,2.5) {};
\node (v42) at (14.5,2.5) {};
\node (v23) at (-3.5,0) {};
\node (v25) at (-1.5,0) {};
\node (v28) at (0.5,0) {};
\node at (2.5,0) {};
\node (v30) at (4.5,0) {};
\node (v34) at (6.5,0) {};
\node (v38) at (8.5,0) {};
\node (v36) at (10.5,0) {};
\node (v40) at (12.5,0) {};
\node (v41) at (14.5,0) {};
\node (v19) at (-5,3) {};
\node (v21) at (-4,4) {};
\node (v22) at (-5,-0.5) {};
\node (v24) at (-4,-1.5) {};
\node (v45) at (16,-0.5) {};
\node (v44) at (16,3) {};
\node (v43) at (15,4) {};
\node (v46) at (15,-1.5) {};
\node (v37) at (8.5,4) {};
\node (v32) at (2.5,-1.5) {};
\node[label=left:\textcolor{blue!50}{$x_8$}] (v11) at (-4.5,1.25) {};
\node (v1) at (-2.5,1.25) {};
\node (v12) at (-1.5,1.25) {};
\node (v13) at (0.5,1.25) {};
\node (v4) at (1.5,1.25) {};
\node (v5) at (3.5,1.25) {};
\node (v14) at (4.5,1.25) {};
\node (v15) at (6.5,1.25) {};
\node (v6) at (7.5,1.25) {};
\node (v9) at (9.5,1.25) {};
\node (v16) at (10.5,1.25) {};
\node (v17) at (12.5,1.25) {};
\node (v7) at (13.5,1.25) {};
\node[label=right:\textcolor{blue!50}{$x_3$}] (v18) at (15.5,1.25) {};
\node[label=above:\textcolor{blue!50}{$x_{10}$}] (v2) at (2.5,4) {};
\node[label=below:\textcolor{blue!50}{$x_5$}] (v10) at (8.5,-1.5) {};
\node[label=below:\textcolor{blue!50}{$x_6$}] (v3) at (-0.5,-1.5) {};
\node[label=above:\textcolor{blue!50}{$x_1$}] (v8) at (11.5,4) {};
\draw[thick,blue!50]  (v11.center) to (v12.center);
\draw[densely dashed,thick,blue!50]  (v12.center) to (v13.center);
\draw[thick,blue!50]  (v13.center) to (v14.center);
\draw[densely dashed,thick,blue!50]  (v14.center) to (v15.center);
\draw[thick,blue!50]  (v15.center) to (v16.center);
\draw[densely dashed,thick,blue!50]  (v16.center) to (v17.center);
\draw[thick,blue!50]  (v17.center) to (v18.center);
\draw[very thick]  (v19.center) to (v20.center);
\draw[very thick]  (v20.center) to (v21.center);
\draw[very thick]  (v22.center) to (v23.center);
\draw[very thick]  (v23.center) to (v24.center);
\draw[very thick]  (v20.center) to (v23.center);
\draw[very thick]  (v23.center) to (v25.center);
\draw[very thick]  (v20.center) to (v26.center);
\draw[very thick]  (v26.center) to (v25.center);
\draw[densely dashed,very thick]  (v26.center) to (v27.center);
\draw[densely dashed,very thick]  (v25.center) to (v28.center);
\draw[very thick]  (v27.center) to (v29.center);
\draw[very thick]  (v28.center) to (v30.center);
\draw[very thick]  (v27.center) to (v28.center);
\draw[very thick]  (v31.center) to (v32.center);
\draw[very thick]  (v29.center) to (v30.center);
\draw[densely dashed,very thick]  (v29.center) to (v33.center);
\draw[densely dashed,very thick]  (v30.center) to (v34.center);
\draw[very thick]  (v34.center) to (v33.center);
\draw[very thick]  (v33.center) to (v35.center);
\draw[very thick]  (v34.center) to (v36.center);
\draw[very thick]  (v37.center) to (v38.center);
\draw[densely dashed,very thick]  (v35.center) to (v39.center);
\draw[densely dashed,very thick]  (v36.center) to (v40.center);
\draw[very thick]  (v35.center) to (v36.center);
\draw[very thick]  (v39.center) to (v40.center);
\draw[very thick]  (v40.center) to (v41.center);
\draw[very thick]  (v41.center) to (v42.center);
\draw[very thick]  (v42.center) to (v39.center);
\draw[very thick]  (v43.center) to (v42.center);
\draw[very thick]  (v42.center) to (v44.center);
\draw[very thick]  (v41.center) to (v45.center);
\draw[very thick]  (v41.center) to (v46.center);
\node (v47) at (-2.5,2.5) {};
\node (v48) at (-2.5,0) {};
\node (v52) at (1.5,0) {};
\node (v49) at (1.5,2.5) {};
\node (v50) at (3.5,2.5) {};
\node (v54) at (3.5,0) {};
\node (v53) at (7.5,0) {};
\node (v51) at (7.5,2.5) {};
\node (v55) at (9.5,2.5) {};
\node (v56) at (9.5,0) {};
\node (v58) at (13.5,0) {};
\node (v57) at (13.5,2.5) {};
\draw[thick,blue!50]  (v1.center) to (v47.center);
\draw[thick,blue!50]  (v47.center) to (v2.center);
\draw[thick,blue!50]  (v1.center) to (v48.center);
\draw[thick,blue!50]  (v48.center) to (v3.center);
\draw[thick,blue!50]  (v49.center) to (v2.center);
\draw[thick,blue!50]  (v2.center) to (v50.center);
\draw[thick,blue!50]  (v2.center) to (v51.center);
\draw[thick,blue!50]  (v49.center) to (v52.center);
\draw[thick,blue!50]  (v51.center) to (v53.center);
\draw[thick,blue!50]  (v50.center) to (v54.center);
\draw[thick,blue!50]  (v3.center) to (v52.center);
\draw[thick,blue!50]  (v55.center) to (v56.center);
\draw[thick,blue!50]  (v57.center) to (v58.center);
\draw[thick,blue!50]  (v54.center) to (v10.center);
\draw[thick,blue!50]  (v53.center) to (v10.center);
\draw[thick,blue!50]  (v10.center) to (v56.center);
\draw[thick,blue!50]  (v10.center) to (v58.center);
\draw[thick,blue!50]  (v55.center) to (v8.center);
\draw[thick,blue!50]  (v8.center) to (v57.center);
\node (v60) at (-0.5,2.5) {};
\node (v59) at (-0.5,0) {};
\node (v61) at (5.5,2.5) {};
\node (v62) at (5.5,0) {};
\node (v63) at (11.5,0) {};
\node (v64) at (11.5,2.5) {};
\draw[densely dashed,thick,blue!50]  (v59.center) to (v3.center);
\draw[densely dashed,thick,blue!50]  (v60.center) to (v2.center);
\draw[densely dashed,thick,blue!50]  (v2.center) to (v61.center);
\draw[densely dashed,thick,blue!50]  (v62.center) to (v10.center);
\draw[densely dashed,thick,blue!50]  (v10.center) to (v63.center);
\draw[densely dashed,thick,blue!50]  (v64.center) to (v8.center);
\draw[densely dashed,thick,blue!50]  (v60.center) edge (v59.center);
\draw[densely dashed,thick,blue!50]  (v61.center) edge (v62.center);
\draw[densely dashed,thick,blue!50]  (v64.center) edge (v63.center);
\draw[fill=blue!50] (v1) circle (2pt);
\draw[fill=blue!50] (-0.5,1.25) circle (2pt);
\draw[fill=blue!50]  (v4) circle (2pt);
\draw[fill=blue!50]  (v5) circle (2pt);
\draw[fill=blue!50] (5.5,1.25) circle (2pt);
\draw[fill=blue!50]  (v6) circle (2pt);
\draw[fill=blue!50]  (v9) circle (2pt);
\draw[fill=blue!50] (11.5,1.25) circle (2pt);
\draw[fill=blue!50]  (v7) circle (2pt);
\node at (-3.25,3.5) {$p_9$};
\node at (-4.5,2.35) {$p_8$};
\node at (-4.5,0.15) {$p_7$};
\node at (-3.25,-1) {$p_6$};
\node at (2,-0.9) {$p_5$};
\node at (9.1,3.3) {$p_{10}$};
\node at (14.25,3.5) {$p_1$};
\node at (14.25,-1) {$p_4$};
\node at (15.5,0.15) {$p_3$};
\node at (15.5,2.35) {$p_2$};
\draw [thick,decoration={brace,mirror,raise=0.5cm},decorate] (-3.4,-2) -- (2.4,-2);
\node at (-0.5,-3.25) {$L$};
\draw [thick,decoration={brace,mirror,raise=0.5cm},decorate] (2.6,-2) -- (8.4,-2);
\node at (5.5,-3.25) {$M$};
\draw [thick,decoration={brace,mirror,raise=0.5cm},decorate] (8.6,-2) -- (14.4,-2);
\node at (11.5,-3.25) {$R$};
\end{tikzpicture} 
\caption{The ten-point staggered elliptic ladder diagrams, which are allowed to involve any number of loops on either side ($L$ and $R$) and in between ($M$) the pair of massless legs associated with $p_5$ and $p_{10}$. The momentum-space diagram is shown in black, while the dual-space diagram is shown in blue.}
\label{fig:LMRdiagram_p}
\end{figure}
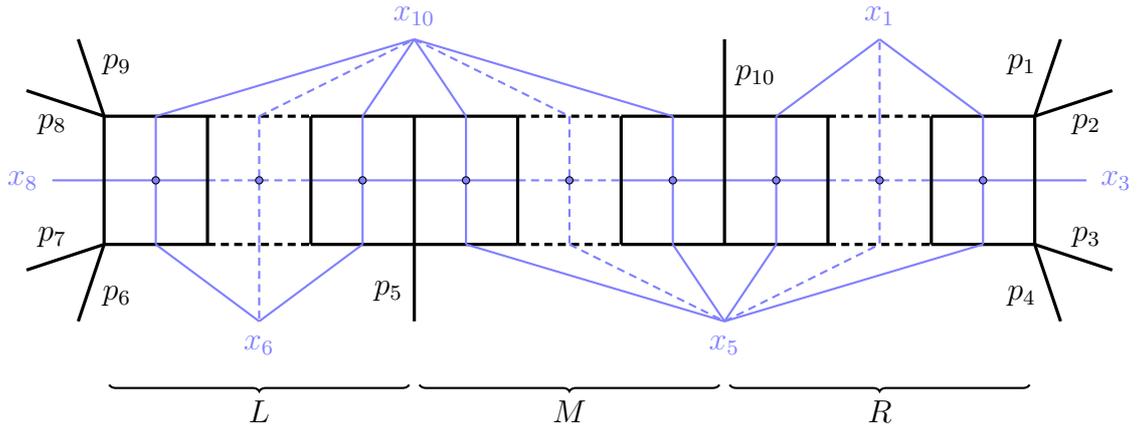

The second family of diagrams we consider are the \emph{staggered elliptic ladder diagrams} depicted in Figure \ref{fig:LMRdiagram_p}, where the dashed lines again indicate any nonzero number of loops. In this case, we allow for the external edges with momentum $p_5$ and $p_{10}$ to connect to different ladder rungs, separated by $M$ loops. However, the staggered ladders still depend on the same seven dual conformal cross-ratios as the unstaggered ones.

In terms of the dual momenta, the corresponding integrals read
\begin{align}
\label{eq: stagg_ladder_integrand}
\mathcal{I}_{L,M,R}=& \, \int \frac{d^4 x_{l_1} \dots d^4 x_{l_{R+M+L}} \qquad x_{3,8}^{2} \,(x_{1,5}^{2})^R \, (x_{5,10}^{2})^M \, (x_{6,10}^{2})^L}{x_{l_1,3}^{2} \Big[ \prod_{i=1}^R x_{l_i, 1}^2 \, x_{l_i, 5}^2 \, x_{l_i, l_{i+1}}^2 \Big] \, \Big[ \prod_{i=R+1}^{R+M} x_{l_i, 10}^2 \, x_{l_i, 5}^2 \, x_{l_i, l_{i+1}}^2 \Big]}\nonumber \\
& \, \times \frac{1}{\Big[ \prod_{i=R+M+1}^{R+M+L} x_{l_i, 10}^2 \, x_{l_i, 6}^2 \, x_{l_i, l_{i+1}}^2 \Big] x_{l_{R+M+L},8}^{2}}\,,
\end{align}
where $x_{l_{R+M+L},l_{R+M+L+1}}^2=1$ for notational compactness. While $\mathcal{I}_{L,0,R}=\mathcal{I}_{L,R}$, we assume $M>0$ and $L \geq R$ in the following. Note that we have picked a particular configuration by disallowing the external edge with momentum $p_{10}$ to be to the left of $p_5$; the other configurations can be obtained from this one via the symmetries $\mathcal{V}$ and $\mathcal{H}$. In total, there are $\lfloor (\ell-1)^2/4 \rfloor$ unique diagrams in this family at $\ell$ loops. 

A Feynman parametrization for the staggered ladders can again be obtained as a soft limit of the one given for traintrack integrals in ref.\ \cite{Bourjaily:2018ycu}:
\begin{equation}
\mathcal{I}_{L,M,R} = \int_0^\infty d^\ell \vec{\alpha} \ d^\ell \vec{\beta} \frac{1}{f_1 \dots f_R  \, f_{R+1} \dots f_{R+M}  \, f_{R+M+1} \dots f_{R+M+L} \, g_\ell} \, ,
\end{equation}
where
{\allowdisplaybreaks
\begin{align}
& f_{k \leq R} = f_{k-1} + \alpha_k + \beta_k + \alpha_k \beta_k + \sum_{j=1}^{k-1} \Big( \alpha_j \beta_k + \alpha_k \beta_j \Big), \\
& f_{R+1} = u_3 f_R + \alpha_{R+1} + \beta_{R+1} + \alpha_{R+1} \beta_{R+1} + \sum_{j=1}^{R} \Big( \alpha_j \beta_{R+1} + u_3 \alpha_{R+1} \beta_j \Big), \\
& f_{R+1<k\leq R+M} = f_{k-1} + \alpha_{k} + \beta_{k} + \alpha_{k} \beta_{k} \nonumber\\
&\qquad \qquad+ \sum_{j=1}^{R} \Big( \alpha_j \beta_{k} + u_3 \alpha_{k} \beta_j \Big) + \sum_{j=R+1}^{k-1} \Big( \alpha_j \beta_{k} + \alpha_{k} \beta_j \Big), \\
& f_{R+M+1} = u_4 u_5 f_{R+M} + \alpha_{R+M+1} + \beta_{R+M+1} + \alpha_{R+M+1} \beta_{R+M+1} \nonumber \\
& \quad \qquad + \sum_{j=1}^{R} \Big( u_4 \alpha_j \beta_{R+M+1} + u_3 \alpha_{R+M+1} \beta_j \Big) + \sum_{j=R+1}^{R+M} \Big( u_4 u_5 \alpha_j \beta_{R+M+1} + \alpha_{R+M+1} \beta_j \Big), \\
& f_{k > R+M+1} = f_{k-1} + \alpha_k + \beta_k + \alpha_k \beta_k + \sum_{j=1}^{R} \Big( u_4 \alpha_j \beta_k + u_3 \alpha_k \beta_j \Big) \nonumber \\
& \qquad \qquad + \sum_{j=R+1}^{R+M} \Big(u_4 u_5 \alpha_j \beta_k + \alpha_k \beta_j \Big) + \sum_{j=R+M+1}^{k-1} \Big(\alpha_j \beta_k + \alpha_k \beta_j \Big), \\
& g_\ell = 1 + \sum_{j=1}^R \Big( v_1 \alpha_j + u_1 \beta_j \Big) + \sum_{j=R+1}^{R+M} \Big( u_2 u_4 u_5 \alpha_j + \frac{u_1}{u_3} \beta_j \Big) + \sum_{j=R+M+1}^\ell \Big( u_2 \alpha_j + v_2 \beta_j \Big),
\end{align}}%
with $f_0=0$. 

While it is less apparent in this case, once again we are able to perform a linear change of variables within the space of $\beta_j$ variables that does not generate quadratic or higher powers in $\beta_k$. In particular, we can apply the change of variables
\begin{subequations}
\label{eq:change_variables_LMR}
\begin{align}
\beta_{1< k \leq R} &\to \beta_k - \beta_{k-1}, \\ \beta_{R+1} &\to \beta_{R+1} - u_3 \beta_R,\\
\beta_{R+1< k \leq R+M} &\to \beta_k - \beta_{k-1}, \\
\beta_{R+M+1} &\to \beta_{R+M+1} - \frac{u_1}{u_3 v_2} \beta_{R+M} - \beta_{R+M+2},\\
\beta_{R+M+1< k < \ell} &\to \beta_k - \beta_{k+1},
\end{align}
\end{subequations}
which renders the staggered ladders linearly reducible in all but one variable, with a single square root in the last integration variable. The corresponding order of integration is $\{ \alpha_1,\dots,\alpha_\ell,\beta_\ell,\dots,\beta_{R+M+2},\beta_1,\dots,\beta_{R+M+1} \}$.

Based both on the leading singularity arguments in appendix~\ref{app: LS} and explicit calculations at low loop orders, we expect these integrals to take the form
\begin{equation}
\label{eq: one-fold for staggered ladders}
\mathcal{I}_{L,M,R} = \int_0^\infty \frac{dx}{(x-z_{+})(x-z_{-})} \, \mathcal{G}^{(2\ell-1)}_{L,M,R}(x,y)\, ,
\end{equation}
where the variables $z_{{\pm}}$ represent the algebraic functions
\begin{equation}
z_{{\pm}}=\frac{u_3 (u_2-u_2 u_4+v_1 -1)-u_1{\pm} \sqrt{(u_1+u_3(u_2-u_2 u_4+v_1 -1))^2-4u_2 u_3^3 u_4 u_5 v_2}}{2 u_3 v_2}.
\end{equation}
We expect $\mathcal{G}^{(2\ell-1)}_{L,M,R}(x,y)$ to be a linear combination of MPLs of uniform transcendental weight $2\ell-1$, and to depend on the same elliptic curve $y$ as appeared in the first family of integrals we studied, which we recall was defined in eq.~\eqref{eq: elliptic curve}.

For the three-loop case $\mathcal{I}_{1,1,1}$, we have explicitly computed $\mathcal{G}^{(2\ell-1)}_{1,1,1}(x,y)$ in terms of MPLs and confirmed that it depends on the elliptic curve $y$. We provide the resulting expression in an ancillary file, accessible at \cite{ancillary_file}. Similar to the case of the elliptic ladder integrals, it would be straightforward to perform the last integration in terms of eMPLs, but we leave that step to future work.

Note that, in contrast to the case of the elliptic ladders $\mathcal{I}_{L,R}$, the final integration kernel in eq.~\eqref{eq: one-fold for staggered ladders} can be brought into a $d$log form via partial fractioning. As a result, $\mathcal{I}_{L,M,R}$ can be written as an integral over a $d$log form. This is also made manifest by the leading singularity analysis in appendix~\ref{app: LS}. While many in the amplitudes field at one point held the expectation that being expressible in terms of $d$log forms guaranteed a function could be written in terms of MPLs, a counterexample to this claim was published in ref.~\cite{Duhr:2020gdd}.
It will be interesting to see whether the staggered ladders $\mathcal{I}_{L,M,R}$ can be expressed in terms of MPLs, or whether they also provide (an infinite number of) further counterexamples.

Finally, let us mention that the staggered ladders satisfy differential equations similar to those of the elliptic ladders:
\begin{equation}
\mathcal{D}_L \,  \mathcal{I}_{L,M,R}= \mathcal{I}_{L-1,M,R}\,,\quad \mathcal{D}_R \,  \mathcal{I}_{L,M,R}= \mathcal{I}_{L,M,R-1}\,,
\label{eq: differential equation staggered}
\end{equation}
where $\mathcal{D}_L$ is precisely the operator defined in eq.~\eqref{eq: differential operator}, and again $\mathcal{D}_R=\mathcal{V}(\mathcal{D}_L)$. In this case, the boundary values $\mathcal{I}_{0,M,R}$ and $\mathcal{I}_{L,M,0}$ are polylogarithmic nine-point ladders, which arise as soft limits of the elliptic ladders $\mathcal{I}_{L,R}$. The two-loop case $\mathcal{I}_{0,1,1}$ was calculated in ref.\ \cite{Wilhelm:2022wow}, while the higher-loop cases can be obtained by taking the soft limit of the integral representation in eq.~\eqref{eq:lblexpression}. Using the change of variables in eq.~\eqref{eq:change_variables_LR}, this integral becomes linearly reducible, and thus easily integrable to MPLs.

\section{Conclusions and Further Directions}
\label{sec:conclusion}

In this work, we considered two families of ten-point ladder diagrams that generalize the elliptic double box. These families differ from each other only by the relative placement of their massless legs: in the first family, two massless legs are attached to opposite sides of the same ladder rung, while in the second family two massless legs are attached to different rungs of the ladder.   
In addition to showing that these diagrams all give rise to the same elliptic curve, we have presented $2\ell$-fold integral representations of both families of diagrams that are linearly reducible in all but one variable. This guarantees that we can carry out the first $2\ell-1$ integrals in terms of multiple polylogarithms, to get expressions of the form
\begin{equation}
\label{eq: one-fold for ladders conclusion}
\drawelliptic = \int_0^\infty \frac{dx}{y} \, \mathcal{G}^{(2\ell-1)}_{L,R}(x,y)\,,
\end{equation}
and 
\begin{equation}
\label{eq: one-fold for staggered ladders conclusion}
\drawstaggered = \int_0^\infty \frac{dx}{(x-z_{+})(x-z_{-})} \, \mathcal{G}^{(2\ell-1)}_{L,M,R}(x,y)\,,
\end{equation}
where $\mathcal{G}^{(2\ell-1)}_{L,R}(x,y)$ and $\mathcal{G}^{(2\ell-1)}_{L,M,R}(x,y)$ are linear combinations of MPLs of weight $2 \ell-1$, and $y$ encodes the same elliptic curve at all loop orders.
 As a direct consequence, we have shown that all these integrals can be expressed in terms of elliptic polylogarithms that depend on \textit{the same} elliptic curve.
 
 While we have argued that the change of variables in eqs.~\eqref{eq:change_variables_LR} and~\eqref{eq:change_variables_LMR} will give rise to integrals that are linearly reducible in all but one variable to all loop orders, our argument falls short of being a proof. Using {\HyperInt}~\cite{Panzer:2014caa}, we have checked this statement through six loops, but at higher loops our expectation is just a conjecture. We hope that a proof can be found for this fact in the future.

We have also derived a set of second-order differential equations that relate the values of these integrals at adjacent loop orders.  In the case of other families of ladder integrals, it has been possible to bootstrap the solutions to analogous differential equations both at particular loop orders~\cite{He:2021non}, and even for finite coupling~\cite{Broadhurst:2010ds,Caron-Huot:2018dsv,He:2020uxy}. It is tempting to attempt a similar feat using the recently developed symbol bootstrap for elliptic Feynman integrals \cite{Morales:2022csr}, and to speculate that there may be a viable finite-coupling description.

The key input needed to bootstrap these integrals at high loop orders would be an understanding of the symbol letters that they depend on. While it is possible to calculate the symbol of the three-loop expressions we have computed, it may be more computationally tractable to instead obtain these symbols by generalizing the symbol integration method~\cite{CaronHuot:2011kk,Li:2021bwg} to the elliptic case, using our one-fold integral expressions in eqs.~\eqref {eq: one-fold for ladders conclusion} and~\eqref{eq: one-fold for staggered ladders conclusion}. 

More broadly, we expect these families of diagrams to provide a useful testing-ground for new elliptic and polylogarithmic technology. They may, for instance, yield further insight into when integrals can be expressed in terms of MPLs, or into the form that might be taken by elliptic scattering amplitudes at finite coupling. Moreover, while we have entirely focused on ladder integrals that give rise to eMPLs in this paper, it seems likely that a similar strategy can be employed to study families of ladder integrals that give rise to integrals over higher-dimensional varieties. In this way, ladder integrals seem to offer a useful laboratory for developing Feynman integral technology in a controlled setting. We leave these interesting questions to future work.

 \subsection*{Acknowledgments}

We thank 
{\"O}mer G{\"u}rdo{\u{g}}an for fruitful discussions,
as well as Song He, Qu Cao and Yichao Tang, for making us aware of their upcoming work \cite{Cao:2023tpx} and kindly coordinating. 
The work of RM, MvH, MW and CZ was supported by the research grant  00025445 from Villum Fonden and the ERC starting grant 757978.

\newpage
\appendix

\section{Leading Singularity of Ladder Integrals}
\label{app: LS}

In this appendix, we will show that the leading singularities of $\mathcal{I}_{L,R}$ and $\mathcal{I}_{L,M,R}$ do not depend on the values of $L$, $M$, or $R$, and hence match the leading singularities of $\mathcal{I}_{1,1}$ and $\mathcal{I}_{1,1,1}$, respectively. This turns out to be a consequence of a more general property of dual-conformally-invariant ladder integrals, namely that one can remove ladder rungs that have no external legs attached to them without changing the leading singularity. Graphically, this corresponds to the rule 
\begin{equation} \label{BCFWbridge}
   \operatorname{LS}  \left( \,
\begin{tikzpicture}[scale=0.35, label distance=-1mm,baseline={([yshift=-.5ex]current bounding box.center)}]
    \draw[thick] (0,0) -- (0,3);
    \draw[thick] (0,0) -- (-3, 0);
    \draw[thick] (0,0) -- (3, 0);
    \draw[thick] (0,3) -- (-3,3);
    \draw[thick] (0,3) -- (3,3);
    
    \draw[thick] (-4.0,3) -- (-7.0,3.4);
    \draw[thick] (-4.0,0) -- (-7.0,-0.4);
    \fill[gray!50] (-4.0,1.5) ellipse (1.5 and 2.1);
    \draw[thick]  (-4.0,1.5) ellipse (1.5 and 2.1);
    \draw[thick] (4.0,3) -- (7.0,3.4);
    \draw[thick] (4.0,0) -- (7.0,-0.4);
    \fill[gray!50] (4.0,1.5) ellipse (1.5 and 2.1);
    \draw[thick]  (4.0,1.5) ellipse (1.5 and 2.1);
    
    \draw[fill=black] (-6.7,1.5) circle (2pt);
    \draw[fill=black] (-6.5,2.3) circle (2pt);
    \draw[fill=black] (-6.5,0.7) circle (2pt);
    
    \draw[fill=black] (6.7,1.5) circle (2pt);
    \draw[fill=black] (6.5,2.3) circle (2pt);
    \draw[fill=black] (6.5,0.7) circle (2pt);
    \node at (0,3.6) {$x_{j}$};
    \node at (0,-0.7) {$x_{i}$};
    
    \node at (-1.3,1.5) {\tiny {$ (AB)$}};
    \node at (1.3,1.5) {\tiny {$ (CD)$}};
        \end{tikzpicture}%
    \, \right) = 
     \operatorname{LS}\left( \,
        \begin{tikzpicture}[scale=0.35, label distance=-1mm,baseline={([yshift=-.5ex]current bounding box.center)}]
            \draw[thick] (0,0) -- (-3, 0);
            \draw[thick] (0,0) -- (2, 0);
            \draw[thick] (0,3) -- (-3,3);
            \draw[thick] (0,3) -- (2,3);
            
            \draw[thick] (-4.0,3) -- (-7.0,3.4);
            \draw[thick] (-4.0,0) -- (-7.0,-0.4);
            \fill[gray!50] (-4.0,1.5) ellipse (1.5 and 2.1);
    			\draw[thick]  (-4.0,1.5) ellipse (1.5 and 2.1);
            \draw[thick] (4.0,3) -- (6.0,3.4);
            \draw[thick] (4.0,0) -- (6.0,-0.4);
            \fill[gray!50] (3.0,1.5) ellipse (1.5 and 2.1);
    			\draw[thick]  (3.0,1.5) ellipse (1.5 and 2.1);
            
            \draw[fill=black] (-6.7,1.5) circle (2pt);
            \draw[fill=black] (-6.5,2.3) circle (2pt);
            \draw[fill=black] (-6.5,0.7) circle (2pt);
            
            \draw[fill=black] (5.7,1.5) circle (2pt);
            \draw[fill=black] (5.5,2.3) circle (2pt);
            \draw[fill=black] (5.5,0.7) circle (2pt);
            
                \end{tikzpicture}%
            \, \right) 
\end{equation}
where the dots on either side of these diagrams can take any graph topology.

To demonstrate property \eqref{BCFWbridge} we work with momentum twistors~\cite{Hodges:2009hk}, which are defined to be 
\begin{equation} 
Z_{i} =    (\lambda_{i}^{\alpha},x_{i}^{\alpha\dot{\alpha}}\lambda_{i\alpha}) 
\end{equation} 
in terms of the usual spinor-helicity variables $p_{i}^{\mu}\sigma_{\mu}^{\alpha\dot{\alpha}}=\lambda_{i}^{\alpha}\tilde{\lambda}_{i}^{\dot{\alpha}}$. Momentum twistors can be combined into an $\mathrm{SL}(4)$-invariant object $\langle i \, j \, k \, l\rangle=\epsilon_{abcd}Z_{i}^{a}Z_{j}^{b}Z_{k}^{c}Z_{l}^{d}$, which is related to the squared differences of dual coordinates we have used in the body of the paper by
\begin{equation}
\frac{\langle i{-}1\,i\,j{-}1\,j \rangle}{\langle i{-1}\,i\rangle\langle j{-1}\,j\rangle}   = x_{i,j}^{2}\,,
\end{equation} 
where $\langle i \, j\rangle=\epsilon_{\alpha\beta}\lambda_{i}^{\alpha}\lambda_{j}^{\beta}$. Each point in dual space corresponds to a line in momentum twistor space; for example, $x_{i}$ corresponds to the line $(i{-}1\,i)$ in momentum twistor space that intersects both $Z_{i-1}$ and $Z_{i}$. Thus, each loop-momentum dual point $x_{l_{i}}$ that is integrated over in eqs.~\eqref{ell_ladder_integrand} and \eqref{eq: stagg_ladder_integrand} can be written as line $(AB)$, $(CD)$, etc. in momentum twistor space. 
The integration measure then becomes
\begin{equation}
    \int dx_{l_{i}} \Leftrightarrow \int \limits_{(AB)} \frac{1}{\langle \lambda_{A} \, \lambda_{B}\rangle^{4}} := \int \frac{d^{4}Z_{A} \, d^{4}Z_{B}}{\langle \lambda_{A} \, \lambda_{B}\rangle^{4}\,\operatorname{Vol}\mathrm{GL}(2)} \,,
\end{equation}
where the factors of $\langle \lambda_{A} \, \lambda_{B}\rangle$ cancel in eqs.~\eqref{ell_ladder_integrand} and \eqref{eq: stagg_ladder_integrand} due to dual conformal invariance (for more details on these types of integrals, see~\cite{ArkaniHamed:2010gh}). 

Using momentum twistors, we can thus write the relevant loop momentum integrals in the diagram on the left side of eq.~\eqref{BCFWbridge} as
\begin{align}
& \int \cdots \frac{d^4 x_{l_L} d^4 x_{l_R} \, (x_{i,j}^2)^2}{x_{l_L,i}^2 \, x_{l_L,j}^2 \, x_{l_L,l_R}^2 \, x_{l_R,i}^2 \, x_{l_R,j}^2} \cdots = \nonumber \\
  & = \int\limits_{(AB),\,(CD)} \cdots \frac{\langle i{-}1\,i\,j{-}1\,j \rangle^2}{\langle AB \,i{-}1\,i\rangle\langle AB \,j{-}1\,j\rangle \langle AB \, CD\rangle\langle CD \,i{-}1\,i\rangle\langle CD \,j{-}1\,j\rangle} \cdots \, .
\label{eq: BCFWbridge integral}
\end{align}
Parametrizing the lines $(AB)$ and $(CD)$ by
\begin{subequations}
    \begin{align}
        Z_{A}&=Z_{i-1} +\alpha_{L}Z_{i} +\epsilon_{L}Z_{j-1}  \,, \quad 
        Z_{B}=Z_{j} +\rho_{L}Z_{i} +\beta_{L}Z_{j-1} \,, \\
        Z_{C}&=Z_{i-1} +\alpha_{R}Z_{i} +\epsilon_{R}Z_{j-1}  \,, \quad 
        Z_{D}=Z_{j} +\rho_{R}Z_{i} +\beta_{R}Z_{j-1} \,,
    \end{align}
\end{subequations}
the integration measure over the line $(AB)$ becomes
\begin{equation} \label{eq: measure_para}
  \int \limits_{(AB)} = \int \frac{d\alpha_{L} \, d\beta_{L} \, d\rho_{L} \, d\epsilon_{L}}{\langle i{-}1\,i\,j{-}1\,j\rangle^{-2}} \,,
\end{equation}
and likewise for the integration measure $\int_{(CD)}$. 
In this parametrization, the denominators involving external points in eq.~\eqref{eq: BCFWbridge integral} read
\begin{align} 
		&\langle AB\,i{-}1\,i\rangle =\epsilon_{L} \, \langle i{-}1\,i\,j{-1}\,j\rangle \,, \qquad 
		&&\langle AB\,j{-}1\,j \rangle = \rho_{L} \, \langle i{-}1\,i\,j{-1}\,j \rangle \,,  \\
  &\langle CD\,i{-}1\,i\rangle =\epsilon_{R} \, \langle i{-}1\,i\,j{-1}\,j\rangle \,, \qquad 
		&&\langle CD\,j{-}1\,j \rangle = \rho_{R} \, \langle i{-}1\,i\,j{-}1\,j \rangle \,.  
	\end{align}
The leading singularity thus involves the residue at $\epsilon_L=\epsilon_R=\rho_L=\rho_R=0$.
On this residue, the remaining propagator in the denominator becomes
\begin{equation}
 \langle AB \, CD\rangle \Big\vert_{\substack{\epsilon_{L}=\epsilon_{R}=0 \\ \rho_{L}=\rho_{R}=0}}=\langle i{-}1\,i\,j{-1}\,j\rangle(\alpha_{L}-\alpha_{R})(\beta_{L}-\beta_{R}) \,,
\end{equation}
which instructs us to take further residues at $\alpha_L=\alpha_R$ and $\beta_L=\beta_R$.
Three factors of $\langle i{-}1\,i\,j{-1}\,j\rangle$ from the residue cancel against two corresponding factors in one of the measures \eqref{eq: measure_para} and one such factor from the numerator of eq.\ \eqref{eq: BCFWbridge integral}.
The combined effect is the same as setting $(AB)=(CD)$ and taking the leading singularity of the diagram without the rung separating $(AB)$ and $(CD)$, which proves eq.\ \eqref{BCFWbridge}.

\section{Action of the Differential Equation on the Ladder Integrand}
\label{app: DEs}

In this appendix, we show that the differential equation in eq.~\eqref{eq: differential equation left} already holds at the level of the Feynman-parameter integral representation in eq.~\eqref{eq:lblexpression}. 

To see this, we first note that $\mathcal{D}_L$ acts only on the variables $u_1, u_2, v_1, v_2$. In the integrand, these variables are only present in the factor $g_\ell$, so this differential operator acts trivially on all other factors. Recalling the particular form of this function,
\begin{equation}
g_\ell = 1 + \sum_{j=1}^R \Big( v_1 \alpha_j + u_1 \beta_j \Big) + \sum_{j=R+1}^\ell \Big( u_2 \alpha_j + v_2 \beta_j \Big)\,,
\end{equation}
we find an additional convenience: the expression is linear in both the cross-ratios and the integration parameters, so each derivative will extract a particular sum of integration variables. Using this, and collecting terms according to their cross-ratio prefactors, we find,
{\allowdisplaybreaks
\begin{align}
\mathcal{D}_L  \frac{1}{g_\ell}  
=\frac{u_2 v_2}{g^3_\ell}\Bigg[& 
2 u_3 u_4 u_5 \left( \sum_{j=1}^R  \alpha_j + \sum_{j=1}^R \beta_j  \right)g_\ell+2  \left( \sum_{j=R+1}^\ell  \alpha_j + \sum_{j=R+1}^\ell \beta_j  \right)g_\ell \nonumber\\
 & + 2 u_3 u_4 u_5 \left( \sum_{j=1}^R  \alpha_j \right)\left( \sum_{j=1}^R \beta_j  \right) + 2 \left( \sum_{j=R+1}^\ell \alpha_j \right)\left( \sum_{j=R+1}^\ell \beta_j \right) \nonumber\\
 & - 2 v_1 u_3 u_4 u_5  \left( \sum_{j=1}^R  \alpha_j + \sum_{j=1}^R \beta_j  \right)\left( \sum_{j=1}^R  \alpha_j \right) \nonumber\\
 & - 2 u_1 u_3 u_4 u_5  \left( \sum_{j=1}^R  \alpha_j + \sum_{j=1}^R \beta_j  \right)\left( \sum_{j=1}^R  \beta_j \right) \nonumber\\
 & - 2 v_2 u_3 u_4 u_5  \left( \sum_{j=1}^R  \alpha_j + \sum_{j=1}^R \beta_j  \right) \left( \sum_{j=R+1}^\ell \beta_j \right) \nonumber\\
 & - 2 u_2 u_3 u_4 u_5  \left( \sum_{j=1}^R  \alpha_j + \sum_{j=1}^R \beta_j  \right) \left( \sum_{j=R+1}^\ell \alpha_j \right) \nonumber\\
 & - 2 v_1 \left( \sum_{j=R+1}^\ell  \alpha_j + \sum_{j=R+1}^\ell \beta_j  \right)\left( \sum_{j=1}^R  \alpha_j \right) \nonumber\\
 & - 2 u_1 \left( \sum_{j=R+1}^\ell  \alpha_j + \sum_{j=R+1}^\ell \beta_j  \right)\left( \sum_{j=1}^R  \beta_j \right) \nonumber\\
 & -2 v_2 \left( \sum_{j=R+1}^\ell  \alpha_j + \sum_{j=R+1}^\ell \beta_j  \right)\left( \sum_{j=R+1}^\ell \beta_j  \right) \nonumber\\
 & -2 u_2 \left( \sum_{j=R+1}^\ell  \alpha_j + \sum_{j=R+1}^\ell \beta_j  \right)\left( \sum_{j=R+1}^\ell \alpha_j  \right) \nonumber\\
 & +2 u_3 \left( \sum_{j=1}^R  \beta_j \right)\left( \sum_{j=R+1}^\ell \alpha_j  \right) + 2 u_4 \left( \sum_{j=1}^R  \alpha_j \right)\left( \sum_{j=R+1}^\ell \beta_j  \right)
 \Bigg]\,.
\end{align}}%
The explicit factors of $g_\ell$ come from the terms that only involve a single derivative. Expanding out these factors, we find that most terms in this expression cancel, and we end up with
\begin{align}
\mathcal{D}_L  \frac{1}{g_\ell}  
=&\, \frac{2 u_2 v_2}{g^3_\ell}\Bigg[ 
u_3 u_4 u_5 \left\{ \sum_{j=1}^R  \alpha_j + \sum_{j=1}^R \beta_j + \left( \sum_{j=1}^R  \alpha_j \right)\left( \sum_{j=1}^R \beta_j  \right) \right\} \nonumber\\
 &\quad\quad\quad +  \sum_{j=R+1}^\ell  \alpha_j + \sum_{j=R+1}^\ell \beta_j + \left( \sum_{j=R+1}^\ell \alpha_j \right)\left( \sum_{j=R+1}^\ell \beta_j  \right) \nonumber\\
 &\quad\quad\quad + u_3 \left( \sum_{j=1}^R  \beta_j \right)\left( \sum_{j=R+1}^\ell \alpha_j  \right) +  u_4 \left( \sum_{j=1}^R  \alpha_j \right)\left( \sum_{j=R+1}^\ell \beta_j  \right)
\Bigg] \nonumber\\
=&\, \frac{2 u_2 v_2 f_{R+L}}{g^3_\ell}\,.
\end{align}
(The last equality may not be immediately obvious from the form of $f_{k}$ presented in eq.~\eqref{eq:bigfdef}, but it can be seen upon expanding the recursion as a sum.) Conveniently, this cancels the factor of $f_{R+L}$ in the denominator of the integrand.

Next, we use the fact that the integrand only depends on the variables $\alpha_l$ and $\beta_l$ in $g_l$, and the function is linear in both. As such, after acting with  $\mathcal{D}_L$ on the integrand it is easy to carry out the integral with respect to both of these variables. We find
\begin{equation}
\int^{\infty}_0 d\alpha_\ell \int^{\infty}_0 d\beta_\ell \, \frac{2 u_2 v_2}{g^3_\ell} = \frac{1}{g_{\ell-1}}\,.
\end{equation}
Thus, all together, we see that
\begin{equation}
\hspace{-0.2cm}\int^{\infty}_0 d\alpha_\ell \int^{\infty}_0 d\beta_\ell \,  \mathcal{D}_L \, \frac{1}{f_1\ldots f_R  \, f_{R+1}\ldots f_{R+L} \, g_\ell} = \frac{1}{f_1\ldots f_R \,  f_{R+1}\ldots f_{R+L-1} \, g_{\ell-1}}\,,
\end{equation}
making it clear that the action of $\mathcal{D}_L$ reduces the number of loops $L$ by one.

Much like the infinite families investigated in ref.~\cite{Caron-Huot:2018dsv}, the existence of this differential equation allows us to extend our definition of $\mathcal{I}_{L,R}$ to define an integral $\mathcal{I}_{0,R}$, as well as $\mathcal{I}_{0,0}$. The first of these integrals is particularly straightforward to write down: one just takes the representation in eq.~\eqref{eq:lblexpression}, and omits all factors $f_{j}$ with $j > R$, as well as the sums that involve the lower limit $j+R+1$ in $g_\ell$. In this form, it becomes clear that the integral only depends on two cross-ratios, $v_1$ and $u_1$. These are exactly the two cross-ratios of the ordinary ladder diagram \cite{Usyukina:1993ch}, which is what we expect to obtain upon deleting all of the loops on one side. They also match the expressions obtained by instead setting $R=0$, just with the exchange of $u_1, v_1$ with $u_2, v_2$. Finally, acting on the one-loop box with this differential operator gives,
\begin{equation}
\mathcal{D}_L  \frac{1}{f_1 g_1} = \frac{2u_2 v_2}{g_1^3}\,,
\end{equation}
which integrates to one. Thus we have $\mathcal{I}_{0,0}=1$.

\bibliography{reference}
\bibliographystyle{JHEP}

\end{document}